\documentclass[aps,twocolumn]{revtex4}

\usepackage{bm}
\usepackage{graphicx}
\usepackage{amsmath}
\usepackage{amssymb}
\usepackage[dvips]{color}

\begin{document}
\title{Solid-on-solid single block dynamics under mechanical vibration}
\author{F. Giacco$^{1}$ , E. Lippiello$^{1,2}$, M. Pica Ciamarra$^{3}$}
\affiliation{%
$^{1}$Dep. of Environmental Sciences, Second University of Naples, 81100
Caserta, Italy\\
$^{2}$CNISM, National Interuniversity Consortium for the Physical Sciences of Matter, 
Italy\\
$^{3}$CNR-SPIN, Department of Physical Sciences, University of Naples Federico II,
80126 Napoli, Italy
}

\begin{abstract}
The suppression of friction between sliding objects, modulated or enhanced by mechanical vibrations,
is well established. However, the precise conditions of occurrence of these phenomena is not well understood.
Here we address these questions focusing on a simple spring--block model,
which is relevant to investigate friction both at the atomistic as well as the macroscopic scale.
This allows to investigate the influence on friction of the properties of the external drive, 
of the geometry of the surfaces over which the block moves, and of the confining force.
Via numerical simulations and a theoretical study of the equations of motion we identify 
the conditions under which friction is suppressed and/or recovered, and evidence the critical
role played by surface modulations and by the properties of the confining force.
\end{abstract}

\maketitle
\section{Introduction}
\label{sec_intro} 
The frictional force between sliding surfaces is
affected by suitable mechanical vibrations, which may 
facilitate or inhibit their relative motion and the associated stick--slip dynamics~\cite{heuberger,rozman,gao,capozza2009}. 
This effect occurs both at the atomistic as well as at the macroscopic scale,
as the physics responsible for friction is expected to be largely the same~\cite{urbakh}.
Indeed, at the nanoscale vibrations have been experimentally observed 
to play a relevant role~\cite{urbakh,gao, brace, diete72, capozza2009, socoliuc2006},
and are relevant for the design of microscopic devices.
Likewise, at the mesoscopic scale numerical~\cite{aharonov} and experimental~\cite{nasuno,tsai,daltonpre2002,johnson2005,johnson2008} works 
showed that external perturbations may affect the frictional forces of 
sheared particulate systems.
These results are frequently connected
to the geophysical scale~\cite{scholz72, diete79, ruina83} where it is possible that earthquakes, 
a stick--slip frictional instability~\cite{marone98,helmstetter},
may be actually triggered by incoming seismic waves~\cite{Stacey}, acting as perturbations.
This phenomenon is regularly observed in numerical simulations of seismic fault models~\cite{PRL2010,Griffa}.

The reduction of friction due to vibrations can be easily attributed to the induced separation of the sliding surfaces. Accordingly, it is surprising that an
increase of the vibrational intensity may also lead to a recovery of the friction coefficient,
a phenomenon observed in systems of sheared and vibrated 
Lennard-Jones particles at zero temperature~\cite{capozza2009}. 
In particular, Ref.~\cite{capozza2009} identified a range of frequencies where friction is suppressed,
and related this range to the amplitude of oscillation as well as to the system inertia and damping constant.
The decrease of the friction coefficient in sheared particulate systems is qualitatively analogous
to the decrease of the viscosity of sheared particulate suspensions, which is known to occur
when particles order in planes parallel to the shearing direction~\cite{tsai}.
Accordingly, one may suspect that the presence of particles in between the sliding surfaces
is essential to reproduce friction suppression and recovery, a question which has not yet been clarified. 
In addition, the influence of the geometry of the oscillating surface on the effective friction coefficient has not been explicitly investigated. 
This is an important point, considering that surfaces of macroscopic objects are expected to be rough, 
while atomistic surfaces are isopotential surfaces, and have a periodicity dictated by the underlying lattice structure. 

Here we address these questions via the theoretical and numerical study of three variants
of a simple   solid--on--solid model, where a block is pulled horizontally by a spring driven at constant velocity.
In all cases the block moves along a surface which is vibrated along the vertical direction, and the role
of both the amplitude and the frequency of vibration is explored.
In model A the surface is flat and the confining stress is vertical; 
in model B the surface is sinusoidal and the confining stress is vertical;
in model C the surface is sinusoidal and the confining stress is normal to the surface.

In the absence of vibrations, spring block models are characterized by a sliding (fluid) phase,
and by a stick--slip (solid) phase, and the transition between the two can be 
analytically obtained~\cite{Vasconcelos}. In the presence of vibrations, these two phases
occur in different regions of the vibrational amplitude/vibrational frequency diagram,
and depend on model details. In particular, model B results
not to be influenced by the vibrations, and its behavior only depends on the
driving velocity. Conversely, in models A and C we find the expected transition from the stick-slip
to the sliding phase for increasing frequency (or amplitude) of the oscillating plate. 
This transition occurs when the maximum acceleration of the oscillating
plate overcomes gravity (or $P_{l}/m$ in model C, $P_l$ being the confining force).
This is the only transition observed in model A.

In model C, a further increase of the frequency leads to a second transition
whereby the system re--enters the stick--slip phase, analogously to what observed
in Ref.~\cite{capozza2009}. 
This second transition originates from a balance between dissipative
and inertial forces. 

The paper is organized as follows: Section II introduces the equations of motion
of the three spring--block models, and the order parameter used to
study the dynamical properties of the system;
Section III reports on results of numerical simulations obtained by changing the
control parameters; different dynamical behaviors are
discussed and demonstrated through theoretical analysis; 
conclusions and open questions are outlined in Section IV.

\begin{figure}[!!t]

\begin{center}
\includegraphics*[width=6.8cm]{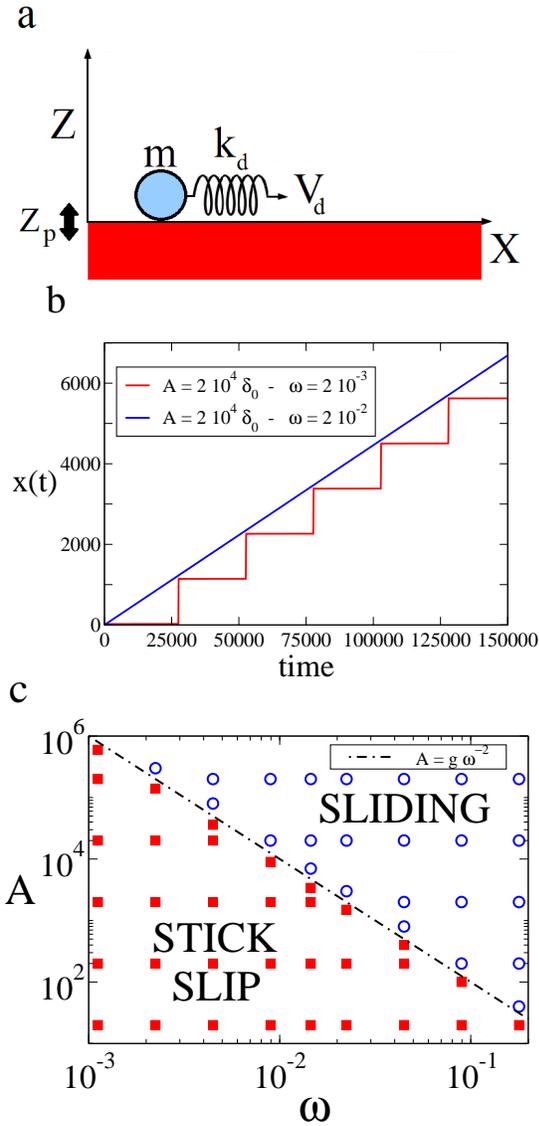}\\
~~\includegraphics*[width=7cm]{fig1b.eps}\\
\includegraphics*[width=7cm]{fig1c.eps}\\
\end{center}
\caption{ \label{fig_modelA} (color online) 
a) Schematic representation of model A: 
a point--particle is driven via an elastic spring on an oscillating flat plane.
b) Time dependence of the horizontal position $x(t)$ of the block for two different oscillation frequencies 
$\omega$ of the plate, at fixed oscillation amplitude $A$, in the sick--slip and in the sliding phase.
c) Phase diagram of model A.  The symbols indicate 
the different phases of the system and are located in positions defined by the values of the parameters 
$A$ and $\omega$ used in the simulations.
A transition from the stick--slip (squares)
to the sliding (circles) phase occurs on increasing $A$ or $\omega$. The transition line (dotted line)
is given by $A=g\omega_{1}^{-2}$.
}
\end{figure}

\section{Single block dynamic}\label{sec-model}

\subsection{Models}
We have investigated three variants of the usual spring--block model, where a point particle of mass $m$ is driven
via a spring of elastic constant $k_d$ on a substrate. One extremity of the driving spring is attached
to the block, while the other moves with an imposed driving velocity $V_d$. 
The block is considered as a point-like body and interacts with an oscillating plate (see Fig. 1a) through contact forces 
which give rise to a viscoelastic response.

We denote with $x$ and $z$ the horizontal and vertical positions of the block, respectively, 
and with $Z_p$ the vertical position of the top surface of the plate. This varies in space and time as 
$Z_p(x,t)=A\sin(\omega t) +A_{x}\sin(\omega_{x}x)$, 
$(A,\omega)$ and $(A_x,\omega_x)$ fixing its temporal and spatial oscillations.

In model A, $A_x = 0$, and the equations of motion along the two directions with respect to a fixed
reference system are given by: 
\begin{equation}\label{equation:uno}
m\ddot{z}=k_{n}(Z_{p}-z)\Theta(Z_{p}-z) -
\gamma_{n}(\dot{z}-\dot{Z_p})\Theta(Z_{p}-z)-mg 
\end{equation}
\begin{equation}\label{equation:due}
m\ddot{x}=-k_{d}(x-V_{\textrm{d}}t) -
\gamma_{t}\dot{x}-k_{t}\mathcal{F}(\dot{x},Z_p-z) 
\end{equation}
where $\Theta$ is the Heaviside step function, $g$ is the gravity, $k_{n,t}$ and $\gamma_{n,t}$ are the
elastic and the damping constants respectively, $k_{d}$ is the elastic constant of the external drive.
The quantity $k_{t}\mathcal{F}(\dot{x})$ is a frictional term which  is only present  when the plate and the block are in contact, 
and this introduces a coupling between the two equations. In particular, following a 
standard procedure to implement an history dependent friction between sheared macroscopic surfaces~\cite{materials},
we assume this term to be proportional to the shear displacement over the lifetime of the contact, which depends on
the vertical motion. If, during the time interval $t$--$t_0$, the block is in contact with the plate,
then the friction term is given by $\mathcal{F}(\dot{x}(t),Z_p-z) =\int_{t_{0}}^{t}\dot{x}(t')dt'$.
The friction term $\mathcal{F}$ is set to zero as soon as the block detaches from the plate.
The Coulomb friction is taken into account through the condition
$|k_{t}\mathcal{F} |< \mu_s N$, 
where $N$ is the resultant of the normal forces and $\mu_s$ is the coefficient of static friction. 
When the Coulomb condition is violated $\mathcal{F}$ is set to zero.

\begin{figure}[!!t]
\begin{center}
\includegraphics[width=9cm]{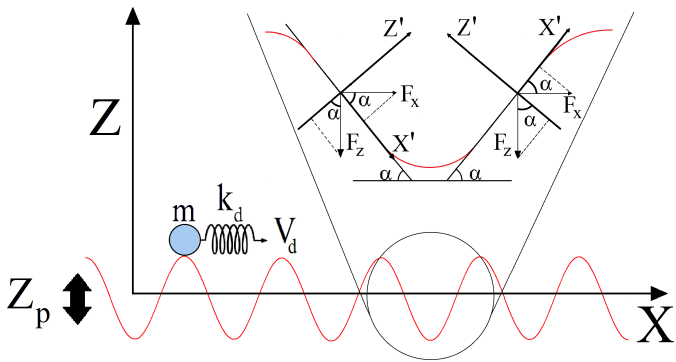}
\end{center}
\caption{ (color online) Models B and C: motion on a  plate with a horizontal periodic corrugation. 
As for the model A, the block is considered as a point-like body interacting with the corrugated plate via
contact forces. The magnification
shows the reference frame used to study the dynamics of the block. In this
frame each force has a vertical and a horizontal component which depends on the
angle $\alpha=\arctan\left(\frac{\partial{Z_{p}}}{\partial{x}}\right)$.
}
\label{figure:corrug_plane}
\end{figure}

In the model B, $A_x > 0$. In this case it is convenient to write the equations of motion in a 
frame of reference which moves along with
the plate, with the horizontal axis tangential to the plate, see Fig.
\ref{figure:corrug_plane}. In this frame the equations of motion are
\begin{eqnarray}
m\ddot{z}' &=& k_{n}(Z_{p}'-z')\Theta(Z_{p}'-z') -
\gamma_{n}(\dot{z}'-\dot{Z_{p}'})\Theta(Z_{p}'-z')+\nonumber \\  &+&
k_{d}(x'-V_{\textrm{d}}t)\sin(\alpha)-mg\cos(\alpha) 
\label{zetaprimo}
\end{eqnarray}
\begin{eqnarray}
m\ddot{x}'&=& -k_{d}(x'-V_{\textrm{d}}t)\cos(\alpha) -
\gamma_{t}\dot{x}'-k_{t}\cdot\mathcal{F}(\dot{x}',Z'_p-z') -\nonumber \\ 
&-&mg\sin(\alpha), 
\label{xprimo}
\end{eqnarray}
where $\alpha=\arctan\left(\frac{\partial{Z_p}}{\partial{x}}\right)$ at fixed
time. 
In the case of friction at atomic level, the periodic corrugation models the
atomic lattice \cite{capozza2009} whereas, in the case of seismic fault, corrugation
represents seismic asperities \cite{asperities}. 

In the model C, we consider the possibility that the mass is not subject to the gravity, but to a confining
force $P_l$ which is always perpendicular to the surface. This is expected to occur in the atomistic systems,
when the block slides over a series of particles periodically arranged on a line, and interacts with them
via Lennard--Jones like potentials, as described in Sec. III C.
In this case, the equations of motion are
\begin{eqnarray}
m\ddot{z}' &=& k_{n}(Z_{p}'-z')\Theta(Z_{p}'-z') -
\gamma_{n}(\dot{z}'-\dot{Z_{p}'})\Theta(Z_{p}'-z')+\nonumber \\  &+&
k_{d}(x'-V_{\textrm{d}}t)\sin(\alpha)-P_{l} 
\label{zetaprimo2}
\end{eqnarray}
and
\begin{eqnarray}
m\ddot{x}'&=&-k_{d}(x'-V_{\textrm{d}}t)\cos(\alpha) -
\gamma_{t}\dot{x}'-k_{t}\mathcal{F}(\dot{x}',Z'_p-z').\;\quad 
\label{xprimo2}
\end{eqnarray}

\subsection{Order parameter}
To differentiate the stick--slip and the flowing phases we introduce an order parameter, defined
as
\begin{equation}
\phi= \frac{\langle (\dot{x}-V_d)^{2} \rangle_{t_a}}{V_d^{2}}
\label{op}
\end{equation}
where the brackets indicate temporal averages over a period $t_a$.
In the flowing phase, the block moves with the external drive velocity
($\dot{x}= V_d$) and therefore 
$\phi=0$. Conversely, in the stick-slip phase, $\phi$ takes a finite value that
depends on the period $t_a$ (or the number of occurred slips during $t_a$).
Indeed, the temporal average in the stick--slip phase includes long stick times
with practically zero block velocity, and short slip intervals with very high block velocity. 
Therefore, the values of $\phi>0$ allow to easily identify the stick-slip phase. 

\subsection{Units and numerical details}
In the following we present results 
obtained by solving via a first order numerical integration the differential equations (1)-(6). 
In the simulations, we have monitored the horizontal and
vertical positions of the block as a function of the control parameters of the plate, $A$  and $\omega$, 
while keeping the values of the other parameters fixed to the typical values used in molecular dynamics simulations~\cite{materials}. 
We choose as numerical units $m=1$, $g=1$ and $k_n=1$. This corresponds to express lengths
in units of $\delta_0=mg/k_n$, i.e.  the rest position of the block inside the plate in the absence of oscillations, and times in units
of $\tau_0=\sqrt{m/k_n}$. The other parameters are fixed to the values 
$k_{t}=0.28$, $\gamma_{n}=7\cdot10^{-2}$, $\gamma_{t}=2.5\cdot10^{-4}$, $\mu_{s}=0.5$,
$k_{d}=5\cdot10^{-4}$ and $V_{d}=2\cdot10^{-5}$, unless otherwise specified.
The integration time step  of the equations of motion is $4\cdot10^{-4}$.
In our simulations the chosen parameters assure a quasi-static regime
($V_d\simeq 0$) and 
stick-slip motion in the limit of no external oscillations ($A=0$, $\omega=0$).
Whereas the details of the dynamics depend on the values of these
parameters, the main results presented below are very robust to changes of the parameters
and the physical properties of the system remain unaltered.

\section{Results}

\subsection{Model A~\label{sec:MA}}
The dynamics in the vertical direction of model A are conveniently investigated in a frame of reference oscillating with the plate,
making the change of variable $y = z-Z_p$. In this frame of reference Eq.~(\ref{equation:uno}) takes the form 
\begin{equation}\label{equation:uno1}\nonumber
m\ddot{y}=-k_{n}y\Theta(-y) -
\gamma_{n}\dot{y}\Theta(-y)-mg+m \ddot{Z_p}.
\end{equation}
We first study the case of a constant perturbation $\ddot{Z_p}=0$. 
In this case the above equation has an equilibrium solution $y=-mg/k_n$ 
indicating that the block is always in contact with the plate at a fixed 
penetration depth. Eq. (\ref{equation:due}) presents a stationary solution, $k_tV_{S}=k_d(V_d-V_{S})$, 
when the block moves with a constant velocity $V_{S}=V_d k_d/(k_t+k_d)$. In this case 
the friction term $\mathcal{F}$ grows linearly in time, $\mathcal{F}=V_{S}(t-t_0)$.  
As soon as  $\mathcal{F}$ reaches the Coulomb threshold value $\mu_s m g$,
the block is no longer stable and a slip starts. The slip ends when the whole elastic energy of the
 driving spring is relaxed ($x=V_d t$).  Accordingly when $V_S \ll V_d$ one observes regular stick-slip. 
The duration $t_{stick}$ of the stick phase can be obtained from 
$ k_t V_S t_{stick}=\mu_s m g$, while the length of each slip is given by
$\Delta x_{slip}=\frac{\mu_{s} mg}{k_{t}V_s}(V_d-V_S)=\mu_{s} mg/k_{d}$. 

We now consider the effect of the external perturbation, $\ddot{Z_p}\neq 0$, distinguishing two cases.
When $\ddot{Z_p}$ is always smaller than $g$, then the block is confined within the plate,
and the dynamics along the $x$ direction are exactly the ones described when $A=0$. 
Also the slip length and stick duration are independent of the perturbation parameters $A$ and $\omega$. 
On the other hand, as soon as $\ddot{Z_p}>g$, i.e. $A \omega_1^2>g$, the block undergoes a vertical positive acceleration, 
detaches from the plate and accelerates also in the horizontal direction, aiming to reach the velocity
of the external drive, $\dot{x}(t)=V_{d}$  (Fig. 1b).
In this case, the dynamics along $z$ consist of a series of 
jumps over the plate. In the range of parameters we have explored the duration of the contact interval, of order $\sqrt{m/k_n}$,
is much smaller than the  time of flight, $2A\omega/g$. 
As a consequence, the friction term $\mathcal{F}$ is always negligible and the block follows the external drive. 

Summarizing, in model A by systematically changing the control parameters
we find two different dynamical regimes, stick--slip and sliding, separated by a dynamic transition
occurring when $A=g\omega_{1}^{-2}$, as illustrated in Fig.~\ref{fig_modelA}c. 
This friction--suppression transition is the only one we observe in this model.

\begin{figure}[!!t]
\begin{center}
\includegraphics*[width=7.5cm]{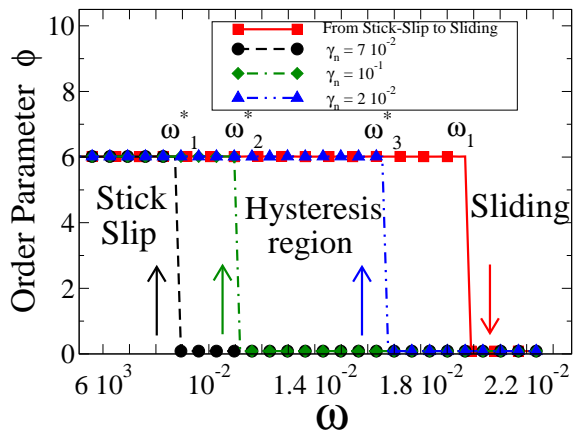}
\end{center}
\caption{\label{fig:hysteresis} (color online) 
Evolution of the order parameter $\phi$ during a frequency cycle performed with
model A. At fixed amplitude we quasi-statically increase the frequency 
until the system transients to the sliding phase at a frequency $\omega_1$ (squares),
which does not depend on $\gamma_n$. The order parameter drops to zero at $\omega_1$. 
On the contrary, the reverse transition from the sliding to the stick--slip phase on decreasing the frequency
occurs at a $\gamma_n$ dependent frequency $\omega^*$. $\omega^*$ approaches $\omega_1$ on
increasing $\gamma_n$, as illustrated.
}
\end{figure}

The transition from the stick--slip to the sliding phase is characterized by 
an hysteresis region, whose extension depends on the damping parameter $\gamma_{n}$,
as illustrated in Fig.~\ref{fig:hysteresis}.
Precisely, the transition from the stick-slip to the sliding
phase always occurs at the same frequency $\omega_1$, while the reverse transition occurs at a $\gamma_{n}$ dependent 
frequency $\omega^{*}(\gamma_{n})$.
On increasing $\gamma_{n}$, $\omega^{*}(\gamma_{n})$ approaches $\omega_1$, and the hysteresis disappears.
An analogous hysteresis effect is observed when the transition is crossed by varying the amplitude of the perturbation.
The origin of this phenomenon is simply related to an excess of energy accumulated in the
sliding phase. A similar phenomenon occurs in magnetic systems~\cite{risken} and frictionless granular systems~\cite{jammzero}.

\subsection{Model B}
\begin{figure}[!!t]
\begin{center}
\includegraphics*[width=7.5cm]{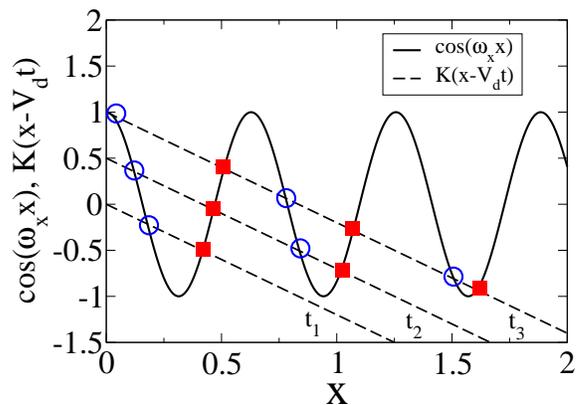}
\end{center}
\caption{ \label{fig:graphsol}(color online) 
Graphical solution of Eq.~(\ref{equation:equili}). Given the values of $K_d$
and $V_{d}$, several solutions may exist as a function of time. In the figure,
$t_{3}>t_{2}>t_{1}$. If the solution is located on the side of the cosine function
with positive slope, then the 
steady-state
position is stable (squares) and the block is stuck. 
Conversely, the other solutions (circles) are unstable.}
\label{effpot}
\end{figure}

We now consider how a periodic modulation of the oscillating surface influences
the dynamics of the system. To this end we fix $A_{x}=10^{3}$ and
$\omega_x=2\cdot10^{-2}$, and consider an external drive with 
$V_d=2\cdot10^{-4}$ and $k_{d}=5\cdot10^{-6}$.
Considering that, in the stick phase, the system is in a 
steady-state
position $x^*$ with 
$\dot{x}(t)=\ddot{x}(t)=0$. We start by identifying the 
steady-state
solutions of Eq.~(\ref{xprimo}),
and to this end we neglect the history dependent frictional term by setting $k_{t}=0$.
The equation for the positions $x^{*}$ is
\begin{equation}\label{equation:equilizero}
-k_{d}(x^{*}-V_{\textrm{d}}t)\cos(\alpha) - mg\sin(\alpha)=0 
\end{equation}
which, combined with the definition of $\alpha$, leads to:
\begin{equation}\label{equation:equili}
\cos(\omega_{x}x^{*})=K(x^{*}-V_{\textrm{d}}t)\quad\textrm{with}\quad
K=\frac{-k_{d}}{A_{x}\omega_{x}mg}.
\end{equation} 
Given the values of $K$ and $V_d$, this equation has several solutions as a function of time. 
This is illustrated in Fig.~\ref{fig:graphsol}, where we show the solutions of Eq.~(\ref{equation:equili}) at three different
times $t_3>t_2>t_1$, indicating with filled squares the stable solutions, and with open circles the unstable ones.

If the system is in a stable solution, then as time increases its position drifts 
with a velocity $V_{\textrm{stick}}\left(x^{*}(t)\right)$
to  different stable  positions, until it enters an unstable solution and starts flowing.
The drift velocity $V_{\textrm{stick}}\left(x^{*}(t)\right)$ can be estimated considering
the response to small perturbations around the 
steady-state
position, i.e. $(x^{*},t)\rightarrow(x^{*}+dx,t+dt)$.
Indeed, from Eq.~(\ref{equation:equili}) we have
\begin{equation}\nonumber
\cos(\omega_{x}(x^{*} +dx ))=K(x^{*} +dx-V_{\textrm{d}}t-V_{\textrm{d}}dt), 
\end{equation}
whose first order expansion leads to
\begin{equation}\nonumber
\cos(\omega_{x}x^{*}) -\omega_{x}\sin(\omega_{x}x^{*})dx =K(x^{*}+dx-V_{\textrm{d}}t-V_{\textrm{d}}dt). 
\end{equation}
The drift velocity is therefore given by
\begin{equation}
V_{\textrm{stick}}(x^{*})=\left.\frac{dx}{dt}\right\vert_{x=x^{*}}=\frac{V_d}{1-\frac{mgA_{x}\omega_{x}^{2}}{k_d}\sin({
\omega_{x}x^{*}})}.
\label{vstick}
\end{equation} 
From Eq.~(\ref{equation:equili}) the stable solutions correspond to $\sin({\omega_{x}x^{*}})<0$ and in particular,
for the investigated parameters $\sin(\omega_{x}x^{*}) \simeq -1$  
and $k_d/mgA_x\omega_x^2 \gg 1$. As a consequence we have 
$V_{\textrm{stick}} \ll V_d \ll 1$ and  this velocity is 
negligible.

When $\cos(\omega_{x}x^*)$ reaches $1$, the subsequent closest solution of Eq.~(\ref{equation:equili})
is no longer stable, and the system starts flowing. Depending on the stored elastic energy
$k_{d}(x(t)-V_{d}t)^{2}$,  one may observe a large slip ($\Delta x^* \ll \omega_{x}^{-1}$), 
or a transition to the sliding phase. 
Accordingly, the horizontal dynamics of model B are surprisingly unaffected by the oscillating perturbation,
and the two regimes are only fixed by the value of the driving velocity (see also Sec.~\ref{sec:dr_vel}).
Conversely, the dynamics along $z$  present a transition from a phase where the block detaches from  the plate ($z'>0$) 
to a phase with $z'<0$.  The same transition also occurs in model C, and we detail the conditions of its occurrence
in the next section.

\subsection{Model C}
In the atomic systems, the modulated surface over which the block slides can be seen as an isopotential surface generated
by a series of atoms periodically arranged on a line. If the block interacts with these atoms via a Lennard--Jones type potential,
then the block is always subject to a force perpendicular to the isopotential surface, which could be attractive or repulsive
depending on the distance of the block from the surface. We mimic this physical scenario with model C,
where we consider the block not to be confined by a
gravitational constant force, but rather by a confining force $P_{l}=1$ which is always perpendicular to the plate at the contact point.

A change in the direction of the confining force, which is the only difference between model B and model C, leads to a very different
dynamics. Indeed, the stability analysis of model B, Eq.s~(\ref{equation:equilizero},~\ref{equation:equili}) and Ref.~\cite{atomicfriction1995},
demonstrates that stick-slip arises even in the no-friction limit ($k_{t}=0$). 
Conversely, the same study performed for model C leads to completely different
results, as in the absence of friction the 
steady-state
equation reduces to
$-k_{d}(x^{*}-V_{\textrm{d}}t)\cos(\alpha)=0$, 
which has no solutions. Therefore, the stick--slip phase is never observed
in the absence of friction.

\begin{figure}[!t]
\begin{center}
 \includegraphics*[width=7.5cm]{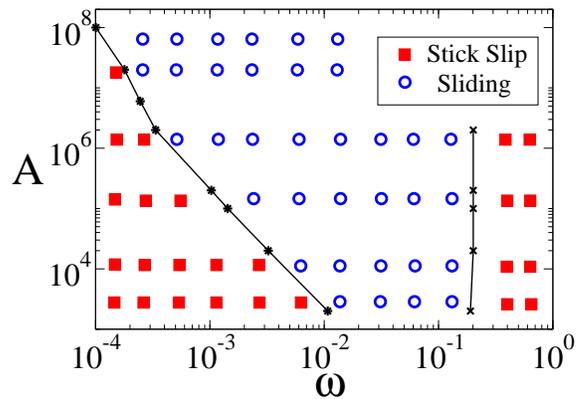}
\end{center}
\caption{ (color online) Phase diagram  of model C.
The points are located according to the values of the parameters $A$ and $\omega$ used
in the simulations.
The system transients from the stick--slip (squares)
to the flowing (circles) phase on increasing the frequency or the amplitude,
analogously to model A (see Fig. 1c).
A second transition from the stick--slip to the flowing
phase is observed at high frequencies.}
\label{figure:doub_trans}
\end{figure}

In the presence of friction, the model exhibits a transition from stick--slip to sliding
when the the plate's acceleration balance the confining term $P_{l}/m$
and the block detaches from the plate. 
As shown in Fig. \ref{figure:doub_trans}, the
modulation of the surface does not strongly affect 
this first transition. Indeed, the stable 
positions during the stick phase
are always located in regions corresponding to angles $\alpha\simeq 0$. 
This leads to a critical curve for the 
transition from stick--slip to sliding phase, $A=P_{l}m^{-1}\omega_{1}^{-2}$, 
which is analogous to that of model A.  

\begin{figure}[!!t]
\begin{center}
\includegraphics*[width=7.5cm]{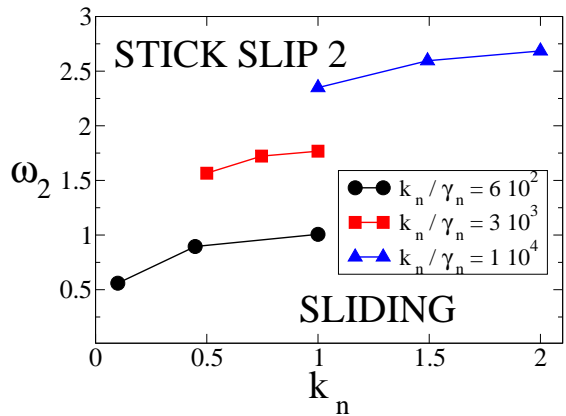}
\end{center}
\caption{~\label{fig:recovery} 
(color online) Dependence of the friction recovery frequency $\omega_{2}$ on the
elastic constant $k_{n}$, for different $k_{n}/\gamma_{n}$ ratios, as indicated.}
\end{figure}

However, this model also exhibits a second transition, in which friction is recovered.
This transition is qualitatively analogous to that observed in Ref.~\cite{capozza2009}.
Indeed, when the oscillating plate's frequency $\omega$ overcomes a critical value $\omega_{2}$, 
the system transients from the sliding to the stick--slip phase, as summarized in Fig.~\ref{figure:doub_trans}.
This transition originates from the balance of the first two terms of Eq.~(\ref{zetaprimo2}), which regulates 
the dynamics of the block along the vertical direction. These two terms 
model elastic and dissipative forces along the direction $z'$, 
and when they cancel out Eq.~(\ref{zetaprimo2})
admits the 
steady-state
solution $k_{d}(x'-V_{\textrm{d}}t)\sin(\alpha)=P_{l}$.  
The balance of these two terms
\begin{eqnarray}
\gamma_{n}(\dot{z}'-\dot{Z_{p}'})\Theta(Z_{p}'-z')\geq k_{n}(Z_{p}'-z')\Theta(Z_{p}'-z'),
\end{eqnarray}
allows to estimate a transition frequency
\begin{equation} 
\omega_{2}\geq\frac{k_{n}}{\gamma_{n}},
\label{omega2}
\end{equation}
that is $A$ independent. This theoretical prediction is supported by the numerical results of Fig.~\ref{fig:recovery}, 
where we show that the recovery frequency $\omega_2$ weakly depends on $k_n$, when computed
at a fixed $k_n/\gamma_n$ ratio.
Contrary to the first transition, this second transition is not
characterized by hysteresis, due to the important role of
the dissipative mechanisms, which dominate over the inertial effects.

We wish to stress that the mechanism leading to the balance of the first two
addends in the rhs of Eq.~(\ref{equation:uno}) does not lead to 
a stable
solution for model A unless $mg=0$. 
This explains why the second transition is not observed in model A,
and clarifies that this transition can only occur if an extra force is introduced 
in the rhs of  Eq.~(\ref{equation:uno}) for the vertical component. 
The balance condition, Eq. (11),  
can be only satisfied in the presence of a modulated interface.
Since there is no qualitative 
difference between model B and model C with respect to the motion along the vertical component $z'$,
this second transition is also observed in the vertical motion of model B.
In both models, for fixed amplitude $A$, the block is embedded in the  plate
when $\omega <\omega_1$ and $\omega>\omega_2$, with $\omega_{2}$ given by Eq.~(\ref{omega2}). 
However, in model B this transition does not affect the 
motion along $x'$ as this is only controlled by the stable solutions of Eq.~(\ref{equation:equili}).
These solutions are not affected by $z'-Z'_p$, as previously discussed.

\subsection{Role of the driving velocity\label{sec:dr_vel}}
Numerical simulations of confined Lennard--Jones particles~\cite{gao} have shown
that the increase of the driving velocity $V_d$ leads the system to the sliding phase,
also in the absence of an external perturbation. Here we show that this transition
is reproduced by the models we have investigated. 

\begin{figure}[!!t]
\begin{center}
\includegraphics*[width=8cm]{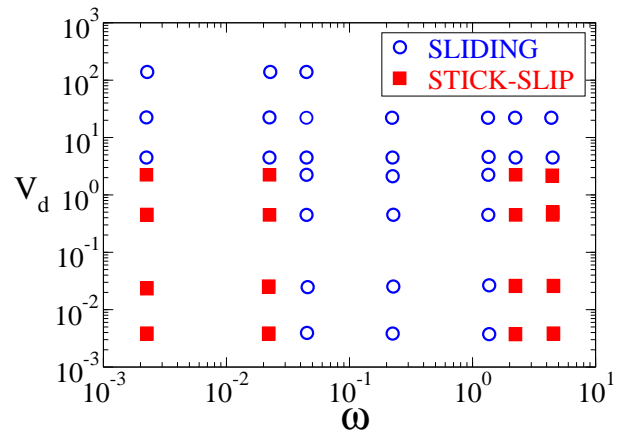}
\end{center}
\caption{\label{fig:model_c_velocity}
(color online) Phase diagram obtained for model C as a function of the driving velocity $V_d$. 
The symbols are located according to the values of the
oscillation frequency $\omega$ and the velocity $V_d$ used in the simulations.
The amplitude of the oscillation is set to the value $A=2\cdot10^{4}$, while
the other model parameters are the same used in the previous analysis of Section III C. }
\end{figure}

In the model A, if the perturbation is turned off,
there are two possible stationary solutions,
depending on the non-linear behavior
of the friction term $\mathcal{F}$. 
One is given by $x(t)=V_S\cdot t$, as described in Section II,
and corresponds to a regime during which the friction term is $\mathcal{F}=V_S(t-t_0)$. 
A second solution occurs when the friction term $k_t\mathcal{F}$ fluctuates so rapidly between its minimum (0)
and its maximum ($\mu_s mg$) values that it can be replaced by its average value ($\mu_s mg/2$).
This is the case when the typical time--scale for friction saturation (of order $\mu_s mg/(k_t V_d)$ ) is much 
smaller than all the other time--scales of the system,
which is certainly the case at very high $V_d$. The corresponding stationary solution
is $x(t)=V_dt-x_0$, with $x_0=\gamma_t \cdot V_d/k_d -\mu_s mg/(2k_d) $.
 The same arguments also apply to the models B and C. As an illustration, we report in 
Fig.~\ref{fig:model_c_velocity} the phase diagram in the $(V_d,\omega)$ plane of model C.
The figure confirms the existence of a transition to the sliding phase for large driving velocity independently
of $\omega$.

\section{Conclusions}
In order to understand the conditions under which mechanical vibrations suppress or enhance the
frictional force between sliding objects, we have considered three variants
of the usual spring--block model, in the presence of an history dependent frictional force.
The models differ for the modulation of the surface over which the block slides, and for
the direction of the confining force. In the model A and C, the equations of motion along the 
two directions are coupled, and on increasing the intensity of the perturbation
we do observe a transition from the stick--slip to the sliding phase. 
Only in the presence of a modulated surface, and of a block confined by a force
which is always normal to this surface, a further increase of the oscillation
frequency leads to a second friction recovery transition, in which the system
transients from the sliding to the stick--slip phase. This transition is analogous
to that observed in Ref.~\cite{capozza2009} in a Lennard--Jones system. Accordingly, our results
clarify that the friction recovery transition is not a peculiarity of many particle systems,
but rather a phenomenon linked to the modulation of the surface over which the block slides.
In the model B mechanical vibrations do not affect the horizontal motion.

Our approach could be adapted to more complex models, suitable for the
study of seismic phenomena, such as the many-blocks spring model. Further works
include the investigation of different contact interactions, and of 
friction laws suggested by recent experimental results~\cite{petri2010}. 

\begin{acknowledgments} 
We gratefully thank professors L. de Arcangelis, C. Godano and Dr. R. Capozza for helpful discussions, and acknowledge 
the financial support of MIUR--FIRB RBFR081IUK (2008) and MIUR--PRIN 20098ZPTW7 (2009).
\end{acknowledgments}


\begin{thebibliography}{99}
\bibitem{heuberger}
M. Heuberger, C. Drummond and J. N. Israelachvili, 
J. Phys. Chem. B {\bf 102}, 5038 (1998).

\bibitem{rozman} M. G. Rozman, M. Urbakh and J. Klafter, 
Phys. Rev. E {\bf 57}, 7340 (1998).

\bibitem{gao} 
J. P. Gao, W.D. Luedtke and U. Landman, 
J. Phys. Chem. B {\bf 102}, 5033 (1998).

\bibitem{capozza2009}
R. Capozza, A. Vanossi, A. Vezzani and S. Zapperi, Phys. Rev. Lett {\bf 103}, 085502 (2009).



\bibitem{urbakh}
M. Urbakh, J. Klafter, D. Gourdon and J. Israelachvili,
Nature  {\bf 430}, 29  (2004). 

\bibitem{brace}
 W. F. Brace and J. D. Byerlee,  Science
{\bf 153}, 990  (1966).

\bibitem{diete72}  J. Dieterich,   J. Geophys.
Res. {\bf 77}, 3690  (1972).



\bibitem{socoliuc2006}
A. Socoliuc, E. Gnecco, S. Maier, O.Pfeiffer, A. Baratoff, R. Bennewitz and 
E. Meyer, 
Science {\bf 313}, 207 (2006).

\bibitem{aharonov} E. Aharonov and D. Sparks, J. Geophys. Res. {\bf109}, B09306
(2004).

\bibitem{nasuno} S. Nasuno, A. Kudrolli and J. P. Gollub, Phys. Rev. Lett.
 {\bf 79}, 949 (1997).
\bibitem{tsai} J. C. Tsai, G. A. Voth and J. P. Gollub, Phys. Rev. Lett. {\bf 91},
064301 (2003).

\bibitem{johnson2005}
P.A. Johnson and X. Jia, Nature {\bf 437}, 871 (2005).

\bibitem{johnson2008} P. A. Johnson, He. Savage, M. Knuth, J. Gomberg
 and C. Marone, Nature {\bf 451}, 57 (2008)

\bibitem{daltonpre2002}
F. Dalton and D. Corcoran, Phys. Rev. E {\bf 63}, 061312
(2001); F. Dalton and D. Corcoran, Phys. Rev. E {\bf 65}, 031310 (2002).

\bibitem{scholz72} C. Scholz, P. Molnar and T. Johnson,  J. Geophys. Res. {\bf 77}, 6392 6 (1972).

\bibitem{diete79} J. Dieterich,   J. Geophys.
Res. {\bf 84}, 2161 (1979).
\bibitem{ruina83} J. R. Rice,  A. L. Ruina,  J. Appl. Mech. {\bf 50}, 343 (1983).


\bibitem{helmstetter} A. Helmstetter and B. E. Shaw, J. Geophys. Res. {\bf 114}, 
B01308 (2009).


\bibitem{marone98} C. Marone, Nature {\bf 391}, 69   1998.

\bibitem{Stacey} 
S. Stacey, J. Gomberg and M. Cocco, J. Geophys. Res. {\bf 110}, B05501 (2005).

\bibitem{PRL2010}
M. Pica Ciamarra, E. Lippiello, C. Godano and  L. de Arcangelis, Phys. Rev. Lett. {\bf 104}, 238001 (2010).

\bibitem{Griffa} 
M. Griffa, E. G. Daub, R. A. Guyer, P. A. Johnson, C. Marone and J. Carmeliet,  Europhys. Lett. {\bf 96}, 14001 (2011).

\bibitem{eplstatistics}
M. Pica Ciamarra, E. Lippiello, L. de Arcangelis and C. Godano, Europhys. Lett. {\bf 95}, 54002 (2011). 



\bibitem{Vasconcelos} G. L. Vasconcelos, Phys. Rev. Lett. {\bf 76}, 4865 (1996). 


\bibitem{asperities}
C.H. Scholz, The mechanics of earthquakes and faulting, Cambridge University Press, 2002.


\bibitem{materials}
P. A. Cundall and O. D. L. Strack, Geotechnique {\bf 29}, 47
(1979).


 
\bibitem{risken} 
H. Risken, \emph{The Fokker-Plank Equation} ( Springer-Verlag,
Berlin, 1984).

\bibitem{jammzero} 
M. Pica Ciamarra and A. Coniglio, Phys. Rev. Lett. {\bf 103}, 235701 (2009).

\bibitem{atomicfriction1995}
T. Gyalog, M. Bammerlin, R. Luthi, E. Meyer and H.
Thomas, Europhys. Lett. {\bf 31}, 269 (1995).

\bibitem{petri2010}
F. Leoni, A. Baldassarri, F. Dalton, A. Petri, G. Pontuale and S. Zapperi, J. Non-Cryst. Solids {\bf 357}, 749 (2011).


\end{thebibliography}
\end{document}